\documentclass[conference,a4paper]{APSIPA2021}
\pdfoutput=1
\usepackage{multirow}
\usepackage{graphicx}
\usepackage{amsmath}
\usepackage[psamsfonts]{amssymb}
\usepackage{amsxtra}
\usepackage{threeparttable}
\usepackage{cite}

\usepackage{booktabs}
\usepackage{multirow}
\usepackage{makecell}
\usepackage{hyperref}
\usepackage[sort&compress, numbers]{natbib}
\usepackage{subcaption}
\usepackage{color}

\begin{document}

\title{Understanding the Tradeoffs in Client-side Privacy for Downstream Speech Tasks}

\author{%
\authorblockN{%
Peter Wu, Paul Pu Liang, Jiatong Shi,\\
Ruslan Salakhutdinov, Shinji Watanabe, Louis-Philippe Morency
}
\authorblockA{%
Carnegie Mellon University, PA, USA \\
E-mail: peterw1@cs.cmu.edu}
}

\maketitle
\thispagestyle{empty}

\begin{abstract}
As users increasingly rely on cloud-based computing services, it is important to ensure that uploaded speech data remains private. Existing solutions rely either on server-side methods or focus on hiding speaker identity. While these approaches reduce certain security concerns, they do not give users client-side control over whether their biometric information is sent to the server. In this paper, we formally define client-side privacy and discuss its three unique technical challenges: (1) direct manipulation of raw data on client devices, (2) adaptability with a broad range of server-side processing models, and (3) low time and space complexity for compatibility with limited-bandwidth devices. Solving these challenges requires new models that achieve high-fidelity reconstruction, privacy preservation of sensitive personal attributes, and efficiency during training and inference. As a step towards client-side privacy for speech recognition, we investigate three techniques spanning signal processing, disentangled representation learning, and adversarial training. Through a series of gender and accent masking tasks, we observe that each method has its unique strengths, but none manage to effectively balance the trade-offs between performance, privacy, and complexity. These insights call for more research in client-side privacy to ensure a safer deployment of cloud-based speech processing services.
\end{abstract}


\section{Introduction}
\label{sec:intro}

Users increasingly rely on cloud-based machine learning models to process their personal data~\cite{chen2020fedhealth,geyer2017differentially,liang2020think,xu2019federated, mcgraw2016personalized, gong2020cloud, anggraini2018speech}.
For example, in cloud-based automatic speech recognition (ASR) systems, audio data recorded on client-side mobile devices are typically uploaded to centralized servers for server-side processing~\cite{leroy2019federated, mcgraw2016personalized, gong2020cloud, anggraini2018speech}, which enables general server-side ASR models to improve over time and enjoy economies of scale. However, there is growing concern that sending raw speech data to the cloud leaves users vulnerable to giving away sensitive personal biometric information such as gender, age, race, and other social constructs~\cite{singh2019profiling}.

In order to have full control over their privacy, users should be able to encrypt their data themselves before uploading to downstream applications. We refer to this type of privacy as \textbf{client-side privacy}, which requires removing sensitive information on the client device while keeping the resulting data compatible with cloud-based services. For example, for a cloud-based ASR service, a client-side privacy algorithm needs to remove biometric information from raw speech data while keeping the resulting audio signal useful for training the ASR model on the cloud. 

As a step towards achieving client-side privacy for speech data, we contribute the following in this work:
\begin{enumerate}
    \item We formally define client-side privacy and describe its three unique technical challenges: (1) direct manipulation and regeneration of raw data on client devices, (2) adaptability with a wide range of server-side processing methods, and (3) low time and space complexity for compatibility with limited-bandwidth client devices.
    \item We study three different client-side privacy approaches for speech: signal processing, disentangled representation learning, and adversarial training.
    \item We conduct experiments on protecting gender and accent information for downstream ASR systems and provide an empirical comparison of current approaches.
\end{enumerate}
We find that each of our three approaches performs well on a subset of metrics, and quantify remaining areas for improvement using multiple privacy metrics. Based on these insights, we propose several extensions for future work and call for more research in client-side privacy to ensure safe cloud-based speech processing.

We proceed by discussing related privacy algorithms in Section~\ref{sec:privacy}. In Section~\ref{sec:client_side_privacy}, we formalize client-side privacy and describe its unique technical challenges. Then, we describe our client-side privacy approaches for downstream ASR in Section \ref{sec:asr_privacy} and detail the experiments in Section \ref{sec:experiments}. Finally, we summarize our results and propose future directions in Section~\ref{sec:conclusion}. All our code and other supplementary material can be found at \url{https://github.com/peter-yh-wu/speech-privacy}.

\section{Related Work}
\label{sec:privacy}

\subsection{Client- and Server-side Privacy}
\label{subsec:client_server_privacy}

One way to view privacy algorithms is by whether they preserve privacy on the client-side, the server-side, or both. Client-side algorithms execute operations on the user's local device, and server-side algorithms run on a remote server \cite{jing1999client, li2020federated, caldas2018leaf}. For example, one way for cloud services to strengthen data privacy is by encrypting on the client-side and decrypting on the server-side \cite{hwang2011serverclient}. For a privacy algorithm to be client-side only, all operations must run on-device without any additional work needed on the server.

\begin{figure}[th]
  \centering
  \includegraphics[width=50mm]{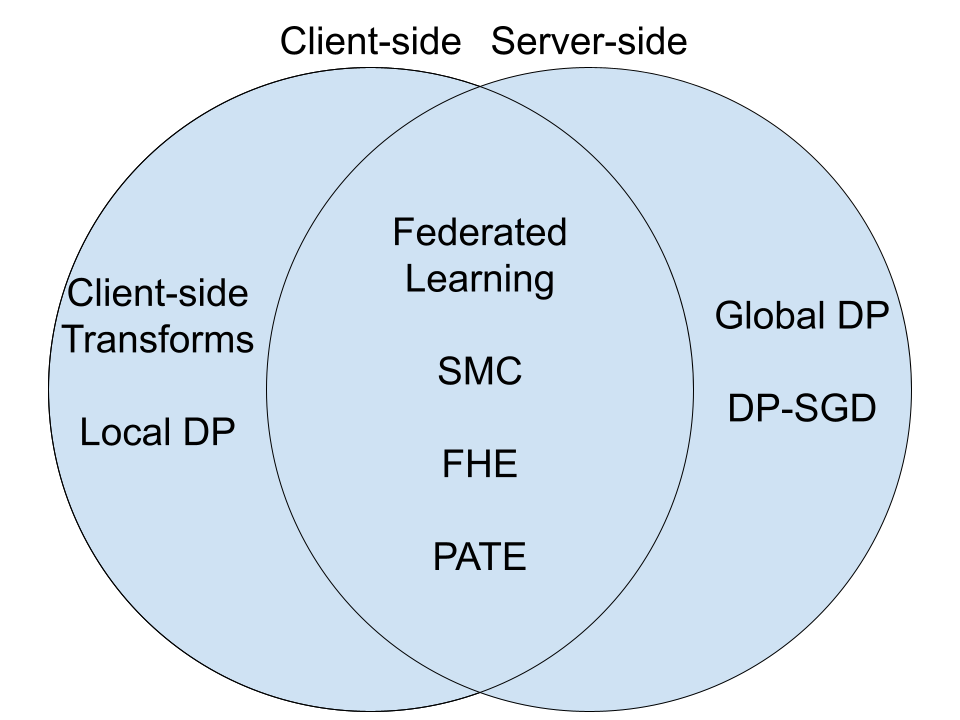}
  \caption{Privacy approaches. Unlike other algorithms, our client-side transforms are housed entirely on the client side and circumvent trade offs inherent in differential privacy.}
  \label{fig:privacy}
\end{figure}

Based on these definitions, we can categorize existing ways to preserve the privacy of data processed by machine learning models. Currently, utilizing cryptography algorithms like secure multi-party computation (SMC) and fully homomorphic encryption (FHE) requires both client and server-side components \cite{zhao2019secure, sun2018private}. Other popular approaches like federated learning and Private Aggregation of Teacher Ensembles (PATE) require both client and server-side modifications as well \cite{li2020federated, caldas2018leaf, papernot2018scalable}. Server-side only approaches also exist, including differentially private SGD (DP-SGD) and global differential privacy \cite{abadi2016deep, de2020overview}. We identify two types of approaches that are client-side only: (1) local differential privacy and (2) methods we refer to as client-side transforms. We define \textbf{client-side transforms} as algorithms that anonymize sensitive information on-device while preserving content needed for downstream tasks. In other words, client-side privacy can be obtained using performant client-side transforms. We note that by being entirely on-device, client-side transforms can be used in conjunction with the aforementioned privacy algorithms by simply applying the client-side transforms first. All three  approaches that we study in this paper are client-side transforms, as detailed in Section \ref{subsec:task_formal}. We observe that client-side transforms are not constrained by trade-offs inherent in local differential privacy. 
Figure \ref{fig:privacy} summarizes the aforementioned categorization of privacy algorithms. 

We note that many client-side transforms exist for downstream tasks simpler than ASR. For example, for the downstream task of storage, client-side encryption is sufficient \cite{wilson2014share}. In contrast, complex downstream tasks like ASR require that the output of the client-side transforms preserve complex content like transcribable audio. Thus, a challenge arises from ensuring that this resulting content also lacks sensitive information like biometrics. Other vulnerabilities like lexical-based ones and sensitive contextual information can also be protected using client-side transforms \cite{silessi2016lexical, reza2017sensitive, caliskan2014privacy}. Since methods like changing language usage can mitigate these vulnerabilities much more easily than biometrics in speech, we focus on client-side privacy for downstream speech tasks here \cite{sun2019mitigating, singh2019profiling}.

\subsection{Privacy-preserving Speech Processing}

ASR models are getting larger and more powerful, making the case for putting them on the server side stronger~\cite{baevski2020wav2vec}. Since the best ASR models are currently on the server side, measures must be taken to ensure user data sent to the server remains private.

Early research on privacy for speech data has focused on voice encryption, which aims to make the original audio hard to recover from the encrypted data \cite{kak1977speech, sridharan1991fast, smaragdis2007smc_asr}. We note that these methods cannot be only on the client side since they require the receiver to decrypt the signal. Another more recent direction focuses on hiding speaker identity~\cite{jin2009speaker, tomashenko2020introducing, aloufi2020privacy, ahmed_preech_privacy_server, ma_fei_privacy_server, nautsh2019speaker, srivastava2020design_vc}. A range of methods relying on server-side operations have been proposed, including rearranging audio segments on the server \cite{ahmed_preech_privacy_server} or leveraging server communication protocols~\cite{ma_fei_privacy_server}. Similar to research in voice encryption, these studies, to our knowledge, do not address the privacy of sensitive information beyond speaker identity like race, gender, or accent.

\textbf{Speaker anonymization} generally refers to approaches that hide speaker identity on the client side \cite{tomashenko2020introducing, srivastava2020design_vc, huang2020sequence_vc_shinji}. Namely, these works leverage voice conversion techniques to transform raw speech into that of another speaker \cite{jin2009speaker, tomashenko2020introducing, nautsh2019speaker,srivastava2020design_vc, huang2020sequence_vc_shinji}. Current approaches are predominantly neural, utilizing adversarial, disentanglement, or other encoder-decoder-related architectures 
\cite{sisman2021vc, zheng2020automatic,  tomashenko2020introducing, nautsh2019speaker,srivastava2020design_vc, noe2020adversarial, ericsson2020adversarial_amnist, aloufi2020privacy, huang2020sequence_vc_shinji}. Since voice conversion has already shown success in anonymizing speaker identity, our client side transforms in this work extend these ideas. Among related work, several address biometric information like gender \cite{noe2020adversarial, ericsson2020adversarial_amnist, chen2007using, kondo2014genderbabble, aloufi2020privacy, stoidis2021protecting}, but to our knowledge only two evaluate on complex downstream tasks like ASR \cite{aloufi2020privacy, stoidis2021protecting}. Since both of these works report high ASR error rates, namely word error rates (WER) above 60\% on LibriSpeech \cite{panayotov2015librispeech}, they are unable to maintain downstream performance while preserving privacy. In our paper, we study three distinct approaches that achieve lower WER while preserving privacy.

We focus on complex downstream tasks like ASR in this work since differential privacy or on-device approaches may be preferable for simpler tasks like classification \cite{ji2014differential, wu2019ordinal, wu2020automatically, liang2020cross}. In Section \ref{sec:experiments}, we also show that differential privacy is not suitable for ensuring privacy in downstream ASR tasks. Additionally, it is much easier to anonymize speaker identity than biometrics like gender, since the former generally requires a much smaller user data distribution shift than the latter \cite{wang2017gender}. Thus, in this paper, we study how well our client-side transforms can anonymize gender and accent, being the first to our knowledge to explore the latter. We note that speaker anonymization is a subtask of our client-side privacy task defined in Section \ref{sec:intro} and detailed below.

\section{Client-side Privacy}
\label{sec:client_side_privacy}

As defined in Section \ref{sec:intro}, client-side privacy refers to privacy obtained only via on-device operations, which remove sensitive information while preserving content needed for downstream tasks. We proceed to formalize the problem statement in Section \ref{subsec:task_formal} and discuss the technical challenges in Section \ref{subsec:technical_challenges}.

\subsection{Problem Statement}
\label{subsec:task_formal}

We start with a set of users $\mathcal{U}$ each of which has access to a client-side device $m_u, u \in \mathcal{U}$. On each client-side device, data $x_u$ is collected which potentially contains information about their private attributes $y_u$ such as gender, age, race, or accent. While it is ideal to leverage shared data collected at a large scale across users, it is also imperative to prevent leakage of private attributes $y_u$ outside of the client's device. Therefore, the goal in client-side privacy is to learn an \textit{encrypted} signal $x_u'$ from $x_u$ using a privacy-preserving function $f_\theta: x_u \rightarrow x_u'$ with parameters $\theta$. $f_\theta$ should perform transformations efficiently and learn an encrypted signal $x_u'$ that balances both fidelity and privacy:

1. \textit{Efficiency}: $|\theta|$ should be small and applying the encoding function $f_\theta: x_u \rightarrow x_u'$ should be fast for cheap inference and storage on resource-constrained mobile devices.

2. \textit{Fidelity}: Given a downstream model trained for a certain task (e.g., ASR) defined on the server, the performance of model on the encrypted signal $x_u'$ should be as close as possible to that of the original signal $x_u$.

3. \textit{Privacy}: One should not be able to decode the private attributes $y_u$ from an encrypted signal $x_u'$ regardless of the function used to predict private attributes.

To avoid confounding factors, both evaluation models (ASR and private attribute classifier) are trained on data completely separate from those used to train encryption approaches. In this paper, we measure fidelity using ASR performance, namely character error rate (CER) and word error rate (WER), and privacy using gender and accent classification accuracy. In other words, low ASR error rate and low classification accuracy would indicate high fidelity and high privacy, respectively. Section~\ref{sec:asr_privacy} contains the efficiency of each of our three approaches, and further details are described in Section~\ref{sec:experiments}.

\subsection{Technical Challenges}
\label{subsec:technical_challenges}


Client-side privacy for downstream speech tasks essentially requires one to \textit{re-generate} raw audio with data-level private attributes masked out. This poses three compelling challenges. First, it requires directly manipulating the user's audio~\cite{srivastava2020design_vc,huang2020sequence_vc_shinji} and re-generating high-dimensional raw speech. Second, the encrypted audio must still be compatible with downstream server tasks without any modification on the server. For downstream tasks like ASR, this means that the encrypted data should still be comprehensible for downstream tasks. This is challenging as it requires preserving information at the high-dimensional data level rather than the feature level. Third, methods for client-side privacy must be efficient and have low time and space complexity to be compatible with limited-bandwidth client devices.

As a result, client-side privacy presents novel challenges over commonly studied server-side methods, particularly on fidelity and efficiency perspectives. 
Furthermore, it is much more challenging to preserve privacy for cluster-level attributes such as race, gender, and accent as compared to individual-level attributes such as speaker identity. This is because transforming data across clusters requires a larger distribution shift than transformations to a new speaker, who could be in the same cluster \cite{wang2017gender}. 

\section{Client-side Transforms for Downstream ASR}
\label{sec:asr_privacy}

Given our definition in Section \ref{sec:privacy}, client-side transforms are one approach to obtaining client-side privacy as defined in Section \ref{sec:intro}. Namely, client-side transforms anonymize sensitive information on-device while preserving content needed for downstream tasks. In this paper, we study three client-side transforms adapted from existing voice conversion literature and analyze their pros and cons.

\subsection{Pitch Standardization}

For our first client-side transform approach, we perform pitch standardization using signal processing \cite{mousa2010voice, laskar2012pitch}. Specifically, we shift the average pitch of each utterance to a predefined value 
while preserving formants. For each utterance, we calculate its fundamental frequency ($F_0$) and then perform a pitch shift from that value to a reference $F_0$. We calculate the sequence of $F_0$'s for each utterance using REAPER,\footnote{\href{https://github.com/google/REAPER}{https://github.com/google/REAPER}} and define the utterance $F_0$ as the average of the non-negative $F_0$'s. We then use the Rubber Band Library to perform pitch shifting with formant preservation using a phase vocoder.\footnote{\href{https://github.com/breakfastquay/rubberband}{https://github.com/breakfastquay/rubberband}} For utterance $u$ with $F_0$ value of $f_u$, we shift its pitch by $12 \log_2 (f_r/f_u)$ semitones, where $f_r$ is the reference $F_0$. This approach easily has the highest efficiency out of our three since it does not depend on a neural model.

\subsection{Disentangled Representation Learning}
\label{disentanglement}

Our second proposed approach uses variational autoencoders (VAE) to disentangle private attributes from non-private speech features~\cite{aloufi2020privacy, stoidis2021protecting, disentangle_vae_avc}. VAEs allow us to learn a set of latent representations that best reconstruct a given input audio signal, while enforcing disentanglement into a set of speaker-dependent and speaker-independent factors \cite{burgess2018understanding}. Here, we define a private attribute encoder $e(x;\theta_p)$ that encodes the input audio signal $x$ into speaker-dependent private factors $z_p$. We also define a content encoder $e(x;\theta_c)$ that encodes $x$ into speaker-independent content factors $z_c$. Since the private factors should capture the private attributes $y$ using a classifier, our goal is to exclude the private factor when decoding the encrypted signal. We optimize the following loss function:
\begin{align}
    \mathcal{L}_{\textrm{dis}} &= \| d(e(x; \theta_c); \theta_d) - x \|_1 - \lambda_p \log P(y|e(x; \theta_c)) \\
    &+ \lambda_{\textrm{dis}} \mathrm{KL} \left( [e(x; \theta_c), e(x; \theta_p)] \ \| \ \mathcal{N}(0, I_d) \right),
\end{align}
where $d(z; \theta_d)$ is a decoder from latent space $z$ back into the audio space. The first term measures the reconstruction of the input signal, the second term measures how well $z_p$ captures the private attributes $y$, and the third term measures disentanglement of $z_c$ and $z_p$ by ensuring minimal correlated entries. $\lambda_p$ and $\lambda_{\textrm{dis}}$ are tunable hyperparameters controlling the tradeoff between privacy disentanglement and performance.

We train a convolutional VAE using the hyperparameters described in Chou et al.~\cite{disentangle_vae_avc}. The model has log-magnitude spectrograms as its input and output acoustic features, and we use the Griffin-Lim algorithm to convert the model output into waveforms~\cite{griffin_lim}. Instance normalization is added to the content encoder $e(x;\theta_c)$ in order to remove speaker information. Also, an adaptive instance normalization layer is added to the decoder in order to add the desired speaker information~\cite{ulyanov2016instance, huang_adaIN}. This allows for the reconstruction of content while transforming speaker information. 
As far as we are aware, this architecture is considered fairly recent among the voice conversion literature \cite{disentangle_vae_avc, huang2020sequence_vc_shinji, sisman2021vc}. Given the success of related architectures in anonymizing speaker identity \cite{yoo2020cyclevaegan, stoidis2021protecting}, we study this model's efficacy in anonymizing biometrics like gender and accent.

\subsection{Adversarial Training}


Reconstruction of high-dimensional signals is difficult and has been shown to cause poor generation quality \cite{dhariwal2020jukebox, rallabandi2019submission, ranzato2014video}. Our final approach attempts to fix this by using adversarial training to ensure high-fidelity generation of encrypted audio \cite{chou_gan, yoo2020cyclevaegan, zhou2021gan}. In the first stage, we disentangle the input audio $x$ into speaker-dependent private factors $z_p$ and speaker-independent content factors $z_c$ learned using an auto-encoder, similarly to Section~\ref{disentanglement}. The second stage trains a generative adversarial network (GAN) \cite{goodfellow2014generative}
to generate realistic audio. The generator is conditioned on the content factor $z_c$ and a new speaker label. 
The discriminator predicts whether an audio sample was from the true dataset or from the generator, and also classifies the speaker. 
The loss function for the generator is
\begin{equation}
    \log c_2(x) + \log (1-c_2(g(x,y)) - \log P_{c_2'}(y|g(x)),
\end{equation}
where $g$ is the generator, and the loss function for the discriminator is
\begin{equation}
    -\log c_2(x) - \log (1-c_2(g(x,y)) - \log P_{c_2'}(y|x).
\end{equation}
This model is suitable for our work because it explicitly separates speaker identity from content twice. We additionally modify this architecture by substituting the speaker label with either the gender label or the accent label. We refer to these three approaches as the speaker, gender, and accent adversarial approaches. While this architecture predates our disentangled one, we observe that our modified methods outperform the disentangled approach on multiple metrics, as detailed in Section \ref{sec:experiments}. We train a convolutional autoencoder using the hyperparameters described in Chou et al.~\cite{chou_gan}. The model has log-magnitude spectrograms as its input and outputs acoustic features, and we use the Griffin-Lim algorithm to convert the model output into waveforms~\cite{griffin_lim}. Thus, we note that our reported results in Section \ref{sec:experiments} can be further improved using more complex vocoders \cite{oord2016wavenet}.

\section{Experiments}
\label{sec:experiments}

Our experiments test whether our proposed approaches are able to balance the trade-offs in fidelity, privacy, and efficiency required for client-side privacy. We test these approaches on masking gender and accents in speech recognition.\footnote{Our code and models are publicly available at \href{https://github.com/peter-yh-wu/speech-privacy}{https://github.com/peter-yh-wu/speech-privacy}.}

\subsection{Setup}

\textbf{Datasets:} We train all of our encryption models on the VCTK corpus, and the ASR model on LibriSpeech~\cite{veaux2016vctk, panayotov2015librispeech}. For both our VCTK and LibriSpeech experiments, we test on speakers unseen during training. We train our privacy attribute classifer on the respective dataset used during testing. In our LibriSpeech experiments, the ASR model and the gender classifier are both evaluated on the test-clean subset. In our VCTK experiments, we evaluate on a hold-out set of 20 speakers comprised of 10 males and 10 females.

\textbf{Classifier:} We use the VGGVox model, a modified version of the VGG-16 CNN, as our privacy attribute classifier~\cite{voxceleb_vggvox, simonyan2014vgg}, slightly modifying the network by adding a ReLU activation followed by a fully-connected layer with size-2 output. We approximate the data available to an adversary by training in two stages: 1. we pre-train the classifier on 100 hours of labeled, unmodified speech from the train data, and 2. we fine-tune the classifier on the encrypted speech of a handful of speakers from the same subset. For each privacy attribute, we measure the masking ability of each encryption approach by calculating the classifier's accuracy on an encrypted version of the test subset after being finetuned on data encrypted using the respective approach.

\textbf{ASR model:} Unless mentioned otherwise, we use a pretrained ESPNet Transformer ASR model to evaluate downstream ASR performance~\cite{watanabe2018espnet}. This model was trained on 960 hours of LibriSpeech data. 


\subsection{Privacy-Fidelity Tradeoff}

Table \ref{tab:tradeoff} compares gender classification accuracy with CER and WER for different levels of Gaussian noise added to the VCTK test data at the waveform level. We increment the standard deviation of the noise by 0.01 for each subsequent experiment. As expected, we observe a negative correlation between classification accuracy and ASR performance. In other words, these results reveal a tradeoff between privacy and fidelity. 

\begin{table}[th]
  \caption{Tradeoff between privacy and fidelity on the VCTK test set for different levels of added Gaussian noise. We observe a negative correlation between classification accuracy and ASR performance, as expected.}
  \label{tab:tradeoff}
  \centering
  \begin{tabular}{ l | c c c}
    \toprule
    \multicolumn{1}{c}{\textbf{Noise}} & \multicolumn{1}{c}{\textbf{Classification}} & \multicolumn{1}{c}{\textbf{CER}} & \multicolumn{1}{c}{\textbf{WER}} \\
    \midrule
    0.00 & $0.99$ & $4.5$ & $9.5$~~               \\
    0.01 & $0.91$ & $11.2$ & $26.7$~~       \\
    0.02 & $0.79$ & $19.3$ & $31.9$~~       \\
    0.03 & $0.68$ & $28.5$ & $41.0$~~       \\
    0.04 & $0.61$ & $33.9$ & $48.1$~~       \\
    \bottomrule
  \end{tabular}
\end{table}

\subsection{Gender Classification}

Table \ref{tab:gender_clf} describes the gender classification accuracy on VCTK and LibriSpeech using gender-masked audio. All proposed approaches can successfully mask gender when no encrypted training examples are available. However, given encrypted training data from a male and female speaker, the signal processing samples become much easier to classify than those generated from the other approaches. This makes sense as pitch shifting would retain some underlying speaker qualities that could be readily identified by a neural classifier. Also, the adversarial approach using the gender-based loss outperforms the speaker-based loss approach here, which reflects how the former explicitly learns to hide gender information.

\begin{table}[th]
  \caption{Gender classification accuracy on two datasets using gender-masked audio. The integer $n$ in each column denotes the number of speakers whose encrypted audio was used to finetune the classifier, where $n/2$ are male and $n/2$ are female. All our voice conversion approaches perform better than the baseline without any masking for $n=0$. For higher $n$ values, the disentanglement approach performs the best, followed by the adversarial approach with the modified gender loss.}
  \label{tab:gender_clf}
  \centering
  \begin{tabular}{ l  r r r r}
    \toprule
    \multicolumn{1}{c}{\textbf{VCTK}} & \multicolumn{1}{c}{\textbf{0}} & \multicolumn{1}{c}{\textbf{2}} & \multicolumn{1}{c}{\textbf{4}} & \multicolumn{1}{c}{\textbf{20}} \\
    \midrule
    No Masking & $0.991$ & $0.991$ & $0.991$ & $0.991$~~             \\
    Signal Processing & $0.590$ & $0.987$ & $0.990$ & $0.997$~~      \\
    Disentanglement & $0.590$ & $0.590$ & $0.598$ & $0.757$~~        \\
    Adversarial (Speaker) & $0.591$ & $0.824$ & $0.840$ & $0.935$~~  \\
    Adversarial (Gender) & $0.590$ & $0.707$ & $0.740$ & $0.877$~~   \\
    \midrule
    \multicolumn{1}{c}{\textbf{LibriSpeech}} &
    \multicolumn{1}{c}{\textbf{0}} & \multicolumn{1}{c}{\textbf{2}} & \multicolumn{1}{c}{\textbf{4}} & \multicolumn{1}{c}{\textbf{20}} \\
    \midrule
    No Masking & $0.972$ & $0.972$ & $0.972$ & $0.972$~~             \\
    Signal Processing & $0.422$ & $0.869$ & $0.887$ & $0.933$~~      \\
    Disentanglement & $0.590$ & $0.627$ & $0.702$ & $0.781$~~        \\
    Adversarial (Speaker) & $0.580$ & $0.712$ & $0.714$ & $0.838$~~ \\
    Adversarial (Gender) & $0.570$ & $0.621$ & $0.628$ & $0.833$~~   \\
    \bottomrule
  \end{tabular}
\end{table}

\subsection{ASR after Gender Encryption}
\label{sec:gender_asr}

Table \ref{tab:gender_asr} describes the ASR results when transcribing gender-encrypted data, measured using mean character and word error rates. For each approach, we provide results on both VCTK and LibriSpeech \cite{veaux2016vctk, panayotov2015librispeech}. All ASR models are pretrained on LibriSpeech as aforementioned. The VCTK model is further finetuned on the data from speakers outside our test set. All finetuned ASR models are tuned on the respective converted data of 20 train speakers. The signal processing approach performs the best, potentially since Griffin-Lim and output distribution priors inherent in neural network architectures introduce artifacts. The adversarial approach using the gender-based loss again outperforms the speaker-based loss approach. This reflects how the former can model less style information than the latter and thus can model more content. For the LibriSpeech dataset, our adversarial method with the modified gender loss performs similarly to the disentangled method in the finetuned scenario. Moreover, our disentanglement and adversarial approaches do not improve when finetuned for the VCTK experiments. This suggests that the ASR model may not be robust enough or our neural converted samples may be acting like adversarial samples during the training procedure \cite{goodfellow2014explaining}. We also note that, compared to Table \ref{tab:tradeoff}, our voice conversion approaches generally achieve lower WER for fixed privacy performances. This suggests that our client-side transforms described in Section \ref{sec:asr_privacy} are more suitable for client-side privacy than other approaches like differential privacy.


\begin{table}[th]
  \caption{ASR performance on two datasets using gender-masked audio. Among our voice conversion approaches, the signal processing method performs the best. For the LibriSpeech dataset, our adversarial method with the modified gender loss performs similarly to the disentangled method in the finetuned scenario. Moreover, our disentenglement and adversarial approaches do not improve when finetuned for the VCTK experiments. This suggests that the ASR model may not be robust enough or our neural converted samples may be acting like adversarial samples during the training procedure \cite{goodfellow2014explaining}.}
  \label{tab:gender_asr}
  \centering
  \begin{tabular}{ l | c c c c}
    \toprule
    \multicolumn{1}{c}{\textbf{}} & \multicolumn{4}{c}{\textbf{VCTK}} \\
    \midrule
    \multicolumn{1}{c}{\textbf{}} & \multicolumn{2}{c}{\textbf{CER}} & \multicolumn{2}{c}{\textbf{WER}} \\
    \multicolumn{1}{c}{\textbf{Method}} & \multicolumn{1}{c}{\textbf{0-Shot}} & \multicolumn{1}{c}{\textbf{Finetune}} &
    \multicolumn{1}{c}{\textbf{0-Shot}} & \multicolumn{1}{c}{\textbf{Finetune}} \\
    \midrule
    No Masking & $4.5$ & $3.4$ & $9.5$ & $4.8$    \\
    Signal Processing & $15.0$ & $7.7$ & $24.7$ & $9.8$         \\
    Disentanglement & $21.1$ & $21.1$ & $35.0$ & $35.0$  \\
    Adversarial (Speaker) & $31.1$ & $31.1$ & $48.5$ & $48.5$     \\
    Adversarial (Gender) & $25.0$ & $25.0$ & $40.1$ & $40.1$ \\
    \bottomrule
  \end{tabular}
  
  \vspace{2mm}
  
  \begin{tabular}{ l | c c c c}
    \toprule
    \multicolumn{1}{c}{\textbf{}} & \multicolumn{4}{c}{\textbf{LibriSpeech}} \\
    \midrule
    \multicolumn{1}{c}{\textbf{}} & \multicolumn{2}{c}{\textbf{CER}} & \multicolumn{2}{c}{\textbf{WER}}  \\
    \multicolumn{1}{c}{\textbf{Method}} & \multicolumn{1}{c}{\textbf{0-Shot}} & \multicolumn{1}{c}{\textbf{Finetune}} &
    \multicolumn{1}{c}{\textbf{0-Shot}} & \multicolumn{1}{c}{\textbf{Finetune}} \\
    \midrule
    No Masking & $2.4$ & $2.4$ & $4.6$ & $4.6$~~             \\
    Signal Processing & $5.0$ & $5.0$ & $8.8$ & $8.8$~~          \\
    Disentanglement & $15.5$ & $15.5$ & $25.0$ & $25.0$~~       \\
    Adversarial (Speaker) & $29.7$ & $17.5$ & $47.5$ & $28.0$~~       \\
    Adversarial (Gender) & $22.0$ & $15.8$ & $36.1$ & $25.3$~~ \\
    \bottomrule
  \end{tabular}
\end{table}

\subsection{Gender Listening Tests}

In addition to our automatic metrics, we perform mean opinion score (MOS) preference tests using eight human listeners. 
Namely, we compare the signal processing, the disentanglement, and the adversarial gender-based loss approaches for the VCTK gender encryption task. For the MOS test, we ask listeners to rate audio samples on a naturalness scale of 1 to 5. 
We use 40 utterances for each test, where 2 are randomly chosen from each test speaker. In other words, each listener listens to 120 unique audio clips. Table \ref{tab:listen} summarizes these results. The signal processing approach performs the best for female speakers and the worst for male speakers. This is likely due to the reference F0 being from a female speaker and the relative absence of artifacts. Also, while the disentanglement approach outperforms the adversarial one in both the classification and ASR metrics, listeners consistently rated the latter higher. This suggests that the disentanglement approach may be standardizing the audio to waveforms that are unnatural to people but suitable for downstream ASR systems. We perceive such utterances as robotic but understandable.

\begin{table}[th]
  \caption{Listening test results on VCTK gender encryption approaches. Our signal processing and adversarial approaches perform the best. While our signal processing approach performs the best for female speakers, our adversarial approach does the best for males, suggesting that the latter is more robust to different speaker attributes.}
  \label{tab:listen}
  \centering
  \begin{tabular}{ l  r r r}
    \toprule
    \multicolumn{1}{c}{\textbf{MOS}} & \multicolumn{1}{c}{\textbf{M}} & \multicolumn{1}{c}{\textbf{F}} & \multicolumn{1}{c}{\textbf{Both}} \\
    \midrule
    Signal Processing & $1.8 \pm 0.2$ & $4.3 \pm 0.2$ & $3.3 \pm 0.2$~~               \\
    Disentanglement & $2.4 \pm 0.5$ & $2.4 \pm 0.5$ & $2.4 \pm 0.5$~~       \\
    Adversarial (Gender) & $2.7 \pm 0.6$ & $3.7 \pm 0.3$ & $3.3 \pm 0.4$~~       \\
    \bottomrule
  \end{tabular}
\end{table}

\subsection{Accent Classification}

Table \ref{tab:accent_clf} describes the accent classification accuracy on VCTK using accent-masked audio. Given that the largest class in the test set contains 31\% of the samples, we observe that both the disentanglement and adversarial approaches are able to successfully fool the accent classifier. Moreover, our adversarial approach using the modified accent loss performs the best, indicating the usefulness of our modified loss function.

\begin{table}[th]
  \caption{Accent classification accuracy using accent-masked audio. Given that the largest class in the test set has 31\% of the samples, our results indicate that both our disentanglement and adversarial approaches successfully fooled the accent classifier. Moreover, our adversarial approach using the modified accent loss performs the best.}
  \label{tab:accent_clf}
  \centering
  \begin{tabular}{ l  r r}
    \toprule
    \multicolumn{1}{c}{\textbf{Classification}} & \multicolumn{1}{c}{\textbf{0}} & \multicolumn{1}{c}{\textbf{20}} \\
    \midrule
    Largest Class & $0.31$ & $0.31$~~               \\
    No Masking & $0.36$ & $0.36$~~               \\
    Disentanglement & $0.29$ & $0.29$~~       \\
    Adversarial (Speaker) & $0.25$ & $0.25$~~       \\
    Adversarial (Gender) & $0.25$ & $0.25$~~       \\
    Adversarial (Accent) & $0.23$ & $0.23$~~       \\
    \bottomrule
  \end{tabular}
\end{table}

\subsection{ASR after Accent Encryption}

Table \ref{tab:accent_asr} describes the ASR results when transcribing gender-encrypted data, measured using mean character and word error rates. Our experimental setup here follows that of the gender ASR experiment. We observe trends similar to Section \ref{sec:gender_asr}. Additionally, ASR results here are consistently better than those in the gender encryption task. This reflects how transforming gender requires a larger data distribution shift than transforming accent.

\begin{table}[th]
  \caption{ASR performance using accent-masked audio. As with our other ASR experiments, we observe that our disentanglement approach outperforms our adversarial one. We also note that these ASR results are consistently better than those for the gender experiment, reflecting the larger data distribution shift required for transforming gender.}
  \label{tab:accent_asr}
  \centering
  \begin{tabular}{ l  r r}
    \toprule
    \multicolumn{1}{c}{\textbf{Speech Recognition}} &
    \multicolumn{1}{c}{\textbf{CER}} & \multicolumn{1}{c}{\textbf{WER}} \\
    \midrule
    Disentanglement & $17.5$ & $29.9$~~       \\
    Adversarial (Speaker) & $26.5$ & $42.4$~~       \\
    Adversarial (Gender) & $19.5$ & $32.4$~~       \\
    Adversarial (Accent) & $23.1$ & $37.4$~~       \\
    \bottomrule
  \end{tabular}
\end{table}

\section{Key Takeaways}

In this section, we outline several key takeaways from our experimental results which we hope will help practitioners working on client-side privacy for complex downstream tasks. 

1. \textit{Pitch standardization} approaches unfortunately are not very effective in keeping gender private, as the gender classification accuracy on data encrypted this way is much higher than those of the neural methods. In other words, this approach yielded artifacts that were readily recognizable by the adversary gender classifier, which implies poorer performance in maintaining privacy. When observing the attention map of the classifier, we noticed that the classifier learned to identify specific patterns that resulted from the pitch shift. Thus, the gender classification accuracy of the neural methods was much lower than those of the signal processing methods.

2. \textit{VAEs and GANs}: VAEs, through use of an encoder, are suitable to learn latent disentangled representations~\cite{higgins2016beta,locatello2019challenging} which are useful in our task of disentangling content from private attributes. 
GANs are also suitable for learning latent representations. While they have been used less in the disentanglement literature, adding attribute-specific loss functions can disentangle sensitive information from content well. While we found GANs to be harder to train than VAEs, GANs that converge appear to perform better.

3. \textit{Memory}: The large differences in memory consumption are a consequence of the large memory costs of using neural models compared to signal processing approaches. Overall, our conclusions point out a ripe opportunity for future work to reconcile the privacy benefits of neural methods with the performance and memory advantages of signal processing approaches.


\section{Conclusion and Future Directions}
\label{sec:conclusion}

In this work, we setup the problem of ensuring the privacy of speech data sent to downstream services that does not rely on any server-side privacy guarantees. We formalized several desirable properties regarding performance, privacy, and computation and performed a large-scale empirical study of existing approaches.
We find that while GAN-based approaches currently have the best tradeoff between gender masking, downstream performance, and memory usage, all existing approaches still fall short of ideal performance.
Our initial empirical analysis opens the door towards more reliable evaluations of the tradeoffs underlying privacy-preserving approaches on the client side, a property crucial for safe real-world deployment of speech systems at scale across mobile devices. In addition to developing privacy-preserving algorithms that satisfy the various desiderata as outlined in this paper, future work should also analyze other downstream speech tasks, including speech translation and other speech recognition settings. 




\bibliographystyle{IEEEtran}
\bibliography{mybib}

\end{document}